\documentclass[aps,prl,twocolumn,reprint,showpacs,preprintnumbers,amsmath,amssymb]{revtex4-1}
\usepackage{graphicx}
\usepackage{dcolumn}
\usepackage{bm}

\usepackage{amsmath}
\usepackage{amssymb}
\usepackage{amsthm}
\usepackage{pinlabel}
\usepackage{natbib}
\usepackage{epstopdf}
\usepackage{color}

\newcommand{\ua}{\uparrow}
\newcommand{\da}{\downarrow}

\usepackage{braket}

\begin{document}

\title{Magnetic-Field-Induced Odd-Frequency Superconductivity in MgB$_2$}

\author{Alex Aperis}\email{alex.aperis@physics.uu.se}
\author{Pablo Maldonado}
\author{Peter M.~Oppeneer}\email{peter.oppeneer@physics.uu.se}
\affiliation{Department of Physics and Astronomy, Uppsala University, P.O.\ Box 516, SE-75120 Uppsala, Sweden}

\vskip 0.7cm
\date{\today}
\begin{abstract}
\noindent 
 In all known superconductors the pairing of fermions is not sensitive to the sign of their time-argument difference, leading to a Cooper pair wavefunction that is even in time/frequency.
Four decades ago it was suggested that odd-frequency superconductivity should in principle be realizable. 
However, observation of odd-frequency superconductivity in bulk materials has remained elusive.
Solving the field-dependent anisotropic Eliashberg equations, we present  \textit{ab initio} evidence for the emergence of odd-frequency pairing under an applied magnetic field in the archetypal superconductor MgB$_2$. We provide the full momentum, frequency and spin resolved dependence and magnetic field-temperature phase diagrams of the even and odd-frequency superconducting pair amplitudes and predict fingerprints of the odd-frequency state in  tunnelling experiments.
\end{abstract}
\date{}
\maketitle

The pair amplitude of the two fermions forming the Cooper pair has to obey the Pauli exclusion principle, which implies that it has to be odd under particle exchange. It is commonly assumed that the pair amplitude is invariant, that is, even, with respect to the sign of the fermionic time-argument difference, which leads to the standard classification of Cooper pair wavefunctions in terms of their symmetry under spatial and spin rotation \cite{Sigrist1991}.
It was however pointed out by Berezinskii \cite{Berezinskii1974new} that the anti-symmetry of the pair amplitude can be fulfilled equally well when it is odd in time/frequency. 
After odd-frequency SC, which cannot be described  within BCS theory, was first proposed 
for $^3$He 
\cite{Berezinskii1974new} it has since then been invoked in several occasions, including disordered Fermi liquids \cite{Kirkpatrick1991}, High-T$_c$ compounds \cite{Balatsky1992}, heavy fermions \cite{Coleman1994} and hydrates \cite{Mazin2005}. 
Odd-frequency SC was also predicted to occur at superconductor to ferromagnet or to normal metal interfaces  where the spatial symmetry is broken \cite{Bergeret2005,Tanaka2007,Eschrig2008,Tanaka2012}. However, in spite of active research and the important ramifications for our understanding of superconductivity unequivocal observation of these exotic states remains elusive \cite{Rochalla2011}.

Remarkably, it was not until recently that the stability of bulk odd-frequency states was theoretically established \cite{Solenov2009,Kusunose2011}, thus opening up the possibility of experimental observations. 
Among all potential states, odd-frequency s-wave spin triplet (OST) SC \cite{Berezinskii1974new} appears as the most robust since it can endure pair breaking by static non-magnetic impurities \cite{Balatsky1992}. However, in a bulk material, OST SC cannot prevail over the even-frequency counterpart unless an extremely strongly coupled and retarded electron-phonon (EP) interaction is at play \cite{Kusunose2011a}. Until a material that fulfills these conditions is found, a promising alternative route is to look for special cases in which odd-frequency pairing could be possible \cite{Fuseya2003,Matsumoto2012,Black-Schaffer2013}. 
The most plausible situation arises when an even frequency s-wave spin singlet (ESS) superconductor is placed in an external magnetic field, since the breaking of time-reversal symmetry is then expected to induce an OST SC component \cite{Matsumoto2012}.

Despite recent progress, a fully microscopic, material-specific theory that supports the reality of OST SC is missing. 
 We have extended the anisotropic Eliashberg framework to include the effect of an external magnetic field, 
 allowing for OST SC as a selfconsistent solution, and show here the unambiguous existence of OST SC in the two-band superconductor MgB$_2$. 
 
The two-band superconductor MgB$_2$ (T$_c$\,=\,39\,K) \cite{Nagamatsu2001,Souma2003} is an ideal candidate to study the magnetic field induced OST SC.
MgB$_2$ is described thoroughly by \textit{ab initio} methods  \cite{An2001}. 
Its Fermi surface consists of sheets with primarily $\pi$ or $\sigma$ orbital character \cite{Kortus2001}. 
High frequency in-plane boron modes lead to an enhanced, very anisotropic EP coupling  that mediates the pairing \cite{Bohnen2001,Choi2002a}. Near the center of the Brillouin zone (BZ) where these modes soften, the EP coupling well exceeds $\lambda_{\bf q}\approx 2$  \cite{Choi2002a} (see also supplementary). Thus, for some regions in momentum space, the EP interaction in MgB$_2$ becomes both strongly retarded and coupled, satisfying the conditions for odd-frequency pairing. 
Due to the inherent anisotropy, SC in MgB$_2$ is characterized by a very anisotropic s-wave, two-gap structure ($\Delta_\pi$\,=\,2.8\,meV and $\Delta_\sigma$\,=\,7\,meV) \cite{Szabo2001} which can be explained with unprecedented precision by fully anisotropic Eliashberg theory \cite{Choi2002,Chen2012}.
Our \textit{ab initio} results for zero field completely confirm the previous calculations, which emphasizes the accuracy of our selfconsistent solutions for finite magnetic fields.

\begin{figure*}[t!]
\includegraphics[width=\textwidth]{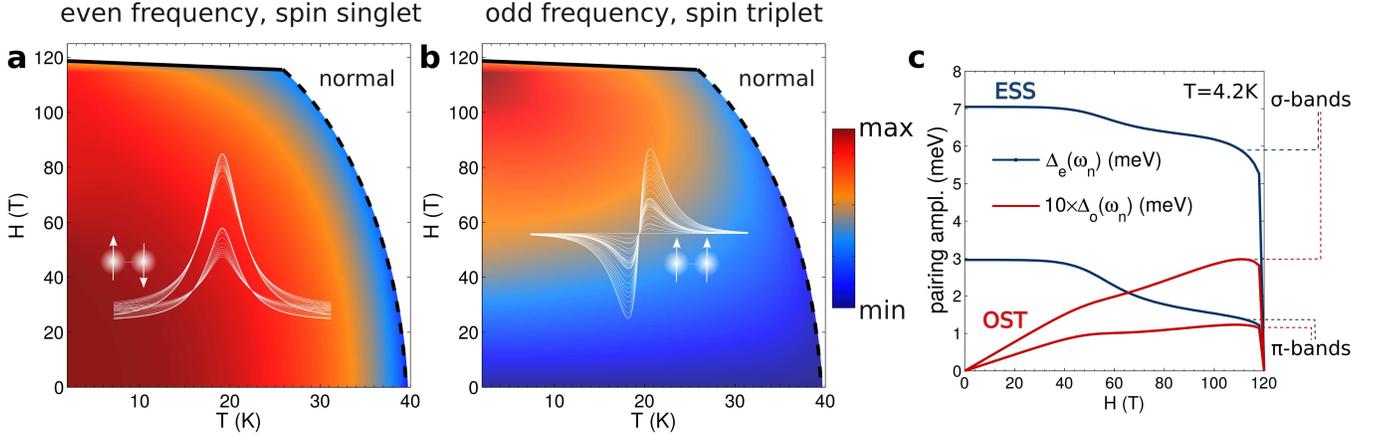}
\caption{\textbf{Temperature and magnetic field dependence of superconductivity in MgB$_2$.} \textbf{a,} H--T phase diagram of the even-frequency superconducting component. Dashed (solid) lines denote second (first) order transitions. \textbf{b,} H--T phase diagram for the odd-frequency superconductivity. The insets show the Matsubara frequency dependence of each component for several magnetic field values. The two gap structure, characteristic of SC in MgB$_2$, can also be discerned. \textbf{c,} Band-resolved magnetic field dependence of the even and odd SC pairing amplitudes at low temperature. The lines correspond to the maximum values in Matsubara space of the momentum averaged superconducting pairing fields on each band, i.e.\ the peaks of the insets in {a}, {b}. Note that the OST pairing amplitude is an order of magnitude smaller than the respective ESS one.}
\label{fig1}
\end{figure*}

In Fig.\ \ref{fig1} we show the calculated magnetic field--temperature phase diagram for the ESS and OST superconducting components. For H\,$>$\,0, the OST SC appears and coexists with the ESS SC in the same H--T regime where the latter is non-zero.
The H--T dependence of the ESS SC follows that of a Pauli limited superconductor; it goes to zero monotonously via a second order phase transition with temperature, except for a very narrow region near the upper critical field (H$_p$), where the transition becomes first order. The transition to the normal state with the field changes from first to second order above T\,$\approx$\,26\,K. 
Our solutions provide the first \textit{ab initio} prediction for the paramagnetic limiting field of MgB$_2$ (see, e.g.\ \cite{Eom2001}) when orbital effects are neglected. 
We find H$_p$=119\,T, which is significantly less than previous estimations ($140$\,T) \cite{Gurevich2007}. At low temperatures, the two ESS SC gaps remain almost constant up to H$_{\pi}\approx39$\,T, as shown in Fig.\ \ref{fig1}\textrm{c}. This is the lowest field strength needed to overcome the binding energy of the Cooper pairs originating from the $\pi$-band. Increasing the field further, leads to partial destruction of the latter gap and the concomitant decrease of the $\sigma$ gap, due to the strong interband EP coupling in this material.

The OST component behaves similar to the ESS as concerns the order of the transitions to the normal state and its temperature evolution.
However, it exhibits a distinctively different magnetic field dependence; it follows a re-entrant behaviour with the field, as is shown in Fig.\ \ref{fig1}\textrm{b}. The trend becomes clearer at higher temperature where the re-entrance peak moves away from  H$_p$. 
From Fig.\ \ref{fig1}\textrm{c} one can observe that there exists, up to H\,$\approx$\,39\,T,  a linear dependence of OST SC on the magnetic field irrespective of the band index. Notably, the OST SC also has a two-gap structure but the pairing is much weaker than the respective ESS.

\begin{figure*}
\includegraphics[width=0.9\textwidth]{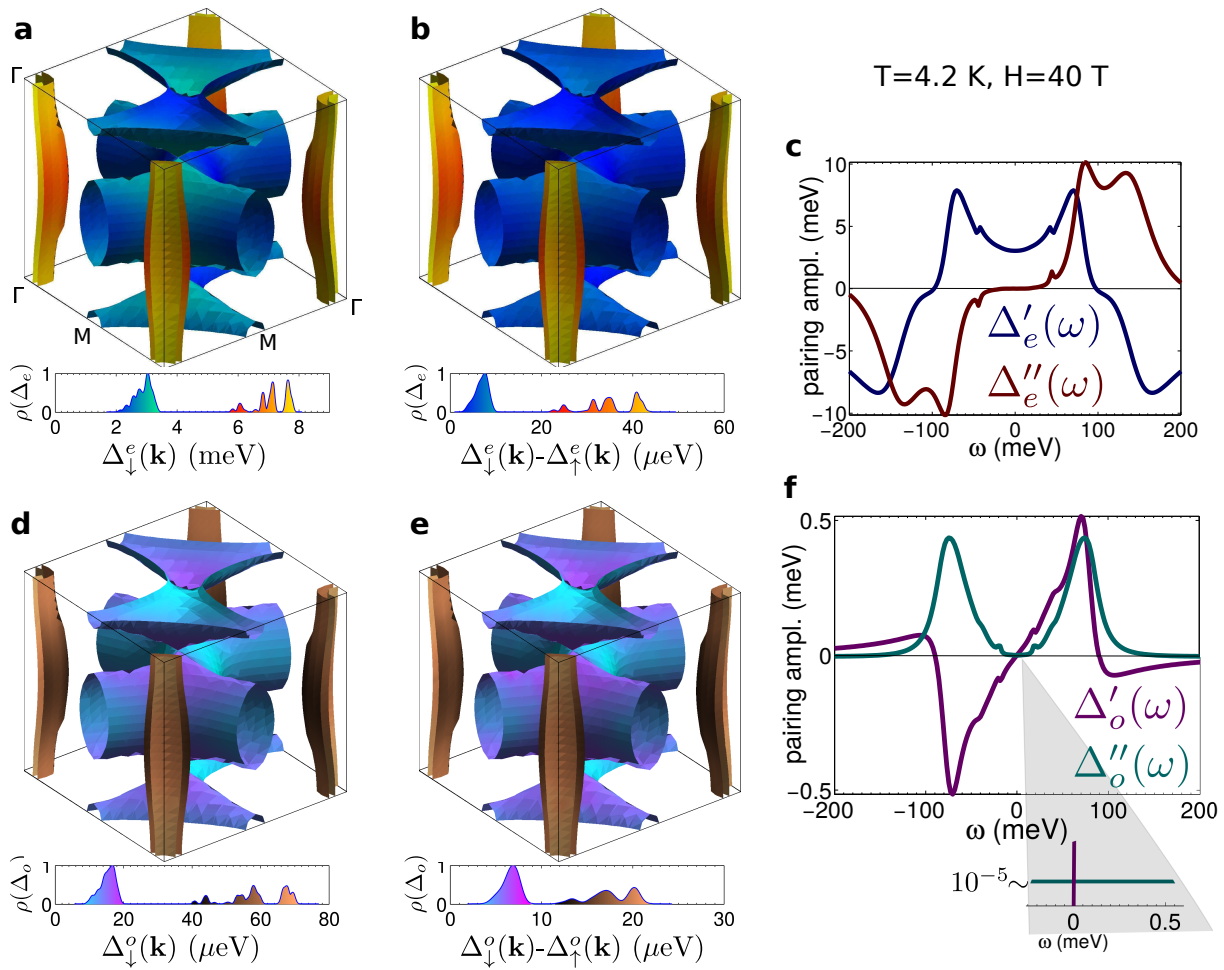}
\caption{\textbf{Momentum and frequency dependence of superconductivity in MgB$_2$ at T\,=\,4.2\,K and H\,=\,40\,T.} \textbf{a,} The Fermi surface of MgB$_2$ in the conventional Brillouin zone coloured by the values of the even-frequency superconducting gap edge for spin-$\da$ quasiparticles. The distribution of the gap edge values is shown in the inset below. \textbf{b,} The difference between spin-$\da$ and spin-$\ua$ even-frequency superconducting gap edges. (\textbf{c,}) Typical frequency dependence of the real ($\Delta_e'(\omega)$) and imaginary ($\Delta_e''(\omega)$) part of the ESS pairing amplitude at a given point on the Fermi surface.
\textbf{d,}\ \textbf{e,} Same as in {a}, {b} but for the odd-frequency superconducting gap edge.
\textbf{f,} Same as in {c} but for the real ($\Delta'_o(\omega)$) and imaginary ($\Delta''_o(\omega)$) parts of the OST pairing amplitude. Near $\omega=0$, $\Delta'_o(\omega)$ increases linearly with frequency whereas $\Delta''_o(\omega)$ is almost constant. The latter is finite at $\omega=0$ as shown in the inset. }
\label{fig2}
\end{figure*}

Next, we examine the Fermi surface momentum dependence of the SC gaps, presented in Fig.\ \ref{fig2}. Since in the Eliashberg framework the superconducting gap retains its full frequency dependence (cf.\ Figs.\ \ref{fig2}\textrm{C},\,\textrm{F}), it is more meaningful to define a superconducting gap edge, instead. In the zero-field case this is given by: $\textrm{Re}\,\Delta_{\bf k}(\omega_{\bf k})=\omega_{\bf k}$ \cite{Choi2002}. However, here the Zeeman effect lifts the degeneracy between spin $\sigma=\ua,\da$ quasiparticles. As a consequence, external probes see a superconducting gap edge that is effectively split in two. 
Moreover, the presence of the OST SC further modifies the gap edge. The relevant expressions can be found from the poles in the system's Green's functions at the Fermi level:
\begin{eqnarray}
\label{gpk}
\!\!\!\!\!
\omega^{\phantom{a}}_{{\bf k},\sigma} = -\sigma{\tilde{\textrm{H}}}({{\bf k},\omega^{\phantom{a}}_{{\bf k},\sigma}}) + \Delta_e({{\bf k},\omega^{\phantom{a}}_{{\bf k},\sigma}}) + \sigma\Delta_o({{\bf k},\omega^{\phantom{a}}_{{\bf k},\sigma}})
\end{eqnarray}
where $\Delta^{e(o)}_{{\bf k},\omega}$ is the order parameter for ESS (OST) SC and $\tilde{\textrm{H}}_{{\bf k},\omega}$ is the renormalized magnetic field term that includes selfenergy contributions (see supplementary). In the above, all quantities are real and the OST spins are perpendicular to H. We define an effective gap edge for each SC component as: $\Delta^{e(o)}_{\ua(\da)}({\bf k})=\Delta_{e(o)}({{\bf k},\omega^{\phantom{a}}_{{\bf k},\ua(\da)}})$.

Since the real part of the OST order parameter vanishes at $\omega$\,=\,0 (Fig.\ \ref{fig2}\textrm{f}), the gap edge of a pure odd-frequency superconductor should be zero. However, due to the presence of the ESS SC and the magnetic field, this is finite in our case. In Fig.\ \ref{fig2}\textrm{a} we show the full momentum dependence of the spin-$\da$ ESS gap edge over the Fermi surface of MgB$_2$. 
The 3D tubular networks are due to the $\pi$-bands, while the almost two dimensional cylinders are due to the $\sigma$-bands  \cite{Kortus2001}. The two-gap structure is similar to the zero-field case \cite{Choi2002} but the momentum anisotropy differs. For the $\pi$-bands we find gap values between $1.8-3.5$ meV, which are close to the zero-field values, while for the $\sigma$-bands we find gap values between $5.6-8$ meV and thus an enhanced anisotropy. 
 Fig.\ \ref{fig2}\textrm{b} shows the difference between the spin-split gap edges of the ESS component. Remarkably, the splitting is more efficient for the $\sigma$-bands where the gap edge is larger. 

As shown in Fig.\ \ref{fig2}\textrm{d}, the OST SC gap edge is very anisotropic and at H\,=\,40\,T it is two orders of magnitude smaller than the respective ESS. Comparing Figs.\ \ref{fig2}\textrm{D} and \ref{fig2}\textrm{B} we find that $\Delta^o_\da({\bf k})$ is proportional to the  difference of the ESS gap edges for the two spin components. This finding directly evidences that the OST superfluid density is proportional to the number density of spin flipped carriers that participate in the ESS Cooper pairs. Therefore, the OST is subordinate to the ESS SC.

\begin{figure*}
\includegraphics[width=0.8\textwidth]{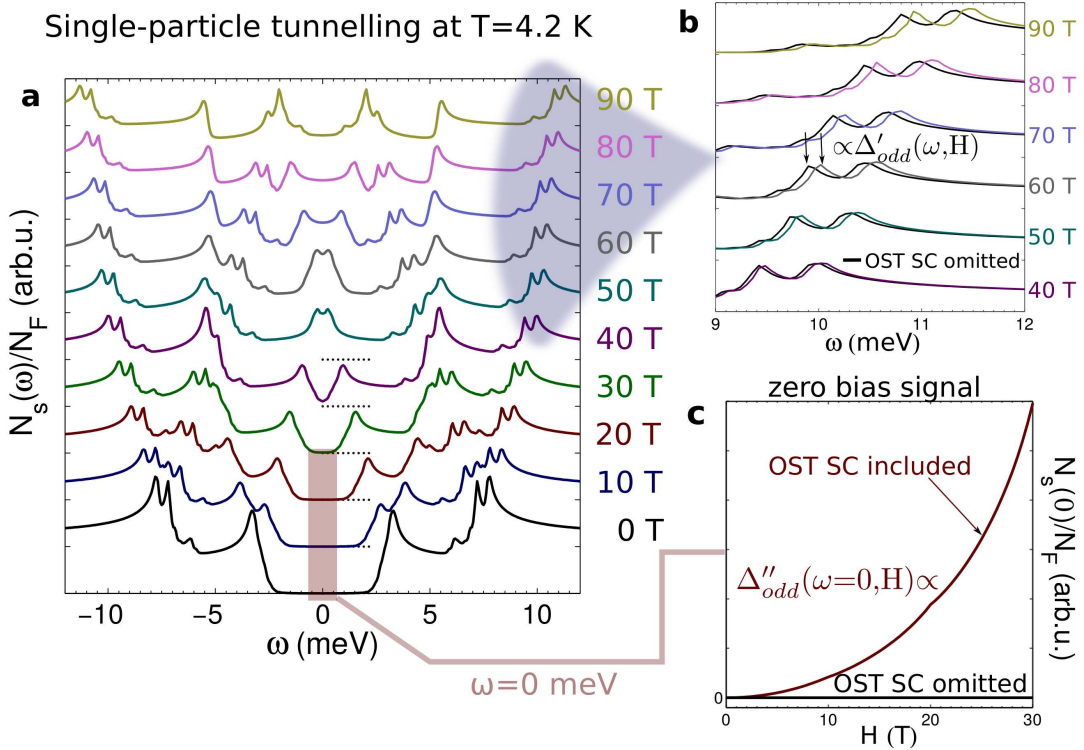}
\caption{\textbf{Dependence of single-particle tunnelling spectra on the magnetic field.} \textbf{a,} Low temperature superconducting density of states of MgB$_2$ for several values of the external magnetic field. \textbf{b,} Comparison between tunnelling spectra when the odd-frequency pairing is included in the Eliashberg calculation and when it is not (black lines). As the field increases, the odd-frequency gap induces a  shift in the tunnelling peaks of the order of 0.1\,meV. This shift is pronounced for the superconducting gap over the $\sigma$-bands. \textbf{c,} In the absence of external bias, we predict a non-zero signal in the tunnelling spectra that is proportional to the imaginary part of the OST SC. This has a distinct magnetic field dependence which detection will serve as definite proof for the existence of odd-frequency SC in MgB$_2$.
\label{fig3}}
\end{figure*}

From the spectral function of our system, we have derived the quasiparticle DOS in the superconducting state, $N(\omega)=\frac{1}{2}\sum_{\sigma}N_\sigma(\omega)$. This quantity is proportional to the differential conductance measured in tunnelling experiments \cite{Chen2012}. The spin-resolved quasiparticle DOS reads
\begin{eqnarray}\label{qpdos}
\!\!\!\frac{N_\sigma(\omega)}{N_F} \!= \textrm{Re}\bigg\langle\frac{|\omega +\sigma \tilde{\textrm{H}}_{{\bf k},\omega}|}{ [ (\omega +\sigma \tilde{\textrm{H}}_{{\bf k},\omega})^2-(\Delta^e_{{\bf k},\omega}+\sigma \Delta^o_{{\bf k},\omega})^2]^{\frac{1}{2}}} \bigg\rangle_{{\! \bf k}}
\end{eqnarray}
where $N_F$ is the DOS at the Fermi level in the normal state and $\langle\ldots\rangle_{\bf k}$ denotes Fermi surface averaging. 
 Using our selfconsistent data in Eq.\ (\ref{qpdos}) we calculate the magnetic field evolution of the tunnelling spectra at low temperature, shown in Fig.\ \ref{fig3}\textrm{a}. For H\,=\,0 we find two peaks around 3\,meV and 7\,meV that signal the quasiparticle excitations above the $\pi$ and the $\sigma$ gaps, respectively \cite{Choi2002,Chen2012}. We also obtain the recently observed fine momentum structure of the latter gap \cite{Chen2012}. As the field increases, the peaks begin to split due to the spin degeneracy lifting. This is clearly visible already at 10\,T for the $\pi$ peak, while at 20\,T the  splitting in both peaks should be very clearly resolved. The gap starts to close around 39\,T where the magnetic field strength is comparable to the minimum gap edge value on the $\pi$-bands.

The odd-frequency component is finite for finite field and increases linearly with $\omega$ in the low frequency regime (cf.\ Fig.\ \ref{fig2}\textrm{f}). Therefore, one would generally expect the OST gap to manifest as a shift of the tunnelling peaks for finite $\omega$, to higher (lower) $\omega$ due to destructive (constructive) interference with the dominant ESS gap. However, due to the smallness of the OST gap, these effects are practically washed out by the dominant ESS component except for two regions. The first is for fields near 39\,T, just before the $\pi$ gap begins to close. There, the OST gap reduces the magnetic field threshold of the zero energy peak by 0.5\,T. The second is at frequencies near the $\sigma$-band peak, where the OST component shifts the peak to the right. For 40\,T the calculated peak shift is around 100\,$\mu$eV and increases with the field as seen in Fig.\ \ref{fig3}\textrm{b}.

We now focus on the zero bias regime, $\omega$\,=\,0. This case is of special interest, since all lifetime effects coming from the EP and the Zeeman interaction are absent. The real part of the induced OST gap is also zero (Fig.\ \ref{fig2}\textrm{f}) but the situation differs from that in a pure OST superconductor since here the ESS component provides a robust gap at the Fermi level. Hence, no low lying excitations should be expected at low temperatures and small, comparing to the $\pi$ gap, magnetic fields. This is certainly true for H\,=\,0 (Fig.\ \ref{qpdos}\textrm{a}).
However, when the field is turned on, the imaginary part of the OST component, $\Delta''_o({\bf k},0)$, becomes finite (Fig.\ \ref{fig2}\textrm{f}). This term effectively contributes an additional lifetime broadening to the quasiparticles and remarkably enforces a nonzero result in Eq.\ (\ref{qpdos}) even at $\omega$\,=\,0. If the OST is not included in the theory, the latter is ideally zero. Therefore, a zero-bias signal in the single-particle tunnelling appears due to the induced OST superconducting component. This signal has a distinct magnetic field dependence, as is shown in Fig.\ \ref{fig3}\textrm{c}. 
This zero-bias plateaux is even more pronounced for T\,$<$\,4.2\,K where the OST SC is more favored (Fig.\ \ref{fig1}\textrm{b}). Thus, we propose that a measurement of the zero-bias tunnelling under magnetic fields less than $\approx$39\,T and at low temperatures would serve as definite proof for the identification of odd-frequency SC in MgB$_2$ and in any superconductor that supports this kind of state. 

Our precise quantitative results predict the existence of OST SC in MgB$_2$, provide an in-depth insight into the microscopic nature of this exotic state and pave the way for the ultimate experimental identification of odd-frequency SC. 
Furthermore, our approach initiates a novel \textit{ab initio} ``roadmap" for the search of such exotic magnetic-field-induced phenomena relevant for both bulk and interface physics \cite{Bergeret2005,Eschrig2008,Tanaka2012}.

\section{Acknowledgments}
We thank G.\ Varelogiannis for fruitful discussions. This work has been supported by the Swedish Research Council (VR) and the Swedish Infrastructure for Computing (SNIC).


\begin{thebibliography}{28}%
\makeatletter
\providecommand \@ifxundefined [1]{%
 \@ifx{#1\undefined}
}%
\providecommand \@ifnum [1]{%
 \ifnum #1\expandafter \@firstoftwo
 \else \expandafter \@secondoftwo
 \fi
}%
\providecommand \@ifx [1]{%
 \ifx #1\expandafter \@firstoftwo
 \else \expandafter \@secondoftwo
 \fi
}%
\providecommand \natexlab [1]{#1}%
\providecommand \enquote  [1]{``#1''}%
\providecommand \bibnamefont  [1]{#1}%
\providecommand \bibfnamefont [1]{#1}%
\providecommand \citenamefont [1]{#1}%
\providecommand \href@noop [0]{\@secondoftwo}%
\providecommand \href [0]{\begingroup \@sanitize@url \@href}%
\providecommand \@href[1]{\@@startlink{#1}\@@href}%
\providecommand \@@href[1]{\endgroup#1\@@endlink}%
\providecommand \@sanitize@url [0]{\catcode `\\12\catcode `\$12\catcode
  `\&12\catcode `\#12\catcode `\^12\catcode `\_12\catcode `\%12\relax}%
\providecommand \@@startlink[1]{}%
\providecommand \@@endlink[0]{}%
\providecommand \url  [0]{\begingroup\@sanitize@url \@url }%
\providecommand \@url [1]{\endgroup\@href {#1}{\urlprefix }}%
\providecommand \urlprefix  [0]{URL }%
\providecommand \Eprint [0]{\href }%
\providecommand \doibase [0]{http://dx.doi.org/}%
\providecommand \selectlanguage [0]{\@gobble}%
\providecommand \bibinfo  [0]{\@secondoftwo}%
\providecommand \bibfield  [0]{\@secondoftwo}%
\providecommand \translation [1]{[#1]}%
\providecommand \BibitemOpen [0]{}%
\providecommand \bibitemStop [0]{}%
\providecommand \bibitemNoStop [0]{.\EOS\space}%
\providecommand \EOS [0]{\spacefactor3000\relax}%
\providecommand \BibitemShut  [1]{\csname bibitem#1\endcsname}%
\let\auto@bib@innerbib\@empty
\bibitem [{\citenamefont {Sigrist}\ and\ \citenamefont
  {Ueda}(1991)}]{Sigrist1991}%
  \BibitemOpen
  \bibfield  {author} {\bibinfo {author} {\bibfnamefont {M.}~\bibnamefont
  {Sigrist}}\ and\ \bibinfo {author} {\bibfnamefont {K.}~\bibnamefont {Ueda}},\
  }\href@noop {} {\bibfield  {journal} {\bibinfo  {journal} {Rev. Mod. Phys.}\
  }\textbf {\bibinfo {volume} {63}},\ \bibinfo {pages} {239} (\bibinfo {year}
  {1991})}\BibitemShut {NoStop}%
\bibitem [{\citenamefont {Berezinskii}(1974)}]{Berezinskii1974new}%
  \BibitemOpen
  \bibfield  {author} {\bibinfo {author} {\bibfnamefont {V.~L.}\ \bibnamefont
  {Berezinskii}},\ }\href@noop {} {\bibfield  {journal} {\bibinfo  {journal}
  {JETP Lett.}\ }\textbf {\bibinfo {volume} {20}},\ \bibinfo {pages} {287}
  (\bibinfo {year} {1974})}\BibitemShut {NoStop}%
\bibitem [{\citenamefont {Kirkpatrick}\ and\ \citenamefont
  {Belitz}(1991)}]{Kirkpatrick1991}%
  \BibitemOpen
  \bibfield  {author} {\bibinfo {author} {\bibfnamefont {T.~R.}\ \bibnamefont
  {Kirkpatrick}}\ and\ \bibinfo {author} {\bibfnamefont {D.}~\bibnamefont
  {Belitz}},\ }\href@noop {} {\bibfield  {journal} {\bibinfo  {journal} {Phys.
  Rev. Lett.}\ }\textbf {\bibinfo {volume} {66}},\ \bibinfo {pages} {1533}
  (\bibinfo {year} {1991})}\BibitemShut {NoStop}%
\bibitem [{\citenamefont {Balatsky}\ and\ \citenamefont
  {Abrahams}(1992)}]{Balatsky1992}%
  \BibitemOpen
  \bibfield  {author} {\bibinfo {author} {\bibfnamefont {A.}~\bibnamefont
  {Balatsky}}\ and\ \bibinfo {author} {\bibfnamefont {E.}~\bibnamefont
  {Abrahams}},\ }\href@noop {} {\bibfield  {journal} {\bibinfo  {journal}
  {Phys. Rev. B}\ }\textbf {\bibinfo {volume} {45}},\ \bibinfo {pages} {13125}
  (\bibinfo {year} {1992})}\BibitemShut {NoStop}%
\bibitem [{\citenamefont {Coleman}\ \emph {et~al.}(1994)\citenamefont
  {Coleman}, \citenamefont {Miranda},\ and\ \citenamefont
  {Tsvelik}}]{Coleman1994}%
  \BibitemOpen
  \bibfield  {author} {\bibinfo {author} {\bibfnamefont {P.}~\bibnamefont
  {Coleman}}, \bibinfo {author} {\bibfnamefont {E.}~\bibnamefont {Miranda}}, \
  and\ \bibinfo {author} {\bibfnamefont {A.}~\bibnamefont {Tsvelik}},\ }\href
  {\doibase 10.1103/PhysRevB.49.8955} {\bibfield  {journal} {\bibinfo
  {journal} {Phys. Rev. B}\ }\textbf {\bibinfo {volume} {49}},\ \bibinfo
  {pages} {8955} (\bibinfo {year} {1994})}\BibitemShut {NoStop}%
\bibitem [{\citenamefont {Mazin}\ and\ \citenamefont
  {Johannes}(2005)}]{Mazin2005}%
  \BibitemOpen
  \bibfield  {author} {\bibinfo {author} {\bibfnamefont {I.~I.}\ \bibnamefont
  {Mazin}}\ and\ \bibinfo {author} {\bibfnamefont {M.~D.}\ \bibnamefont
  {Johannes}},\ }\href {http://dx.doi.org/10.1038/nphys126} {\bibfield
  {journal} {\bibinfo  {journal} {Nat. Phys.}\ }\textbf {\bibinfo {volume}
  {1}},\ \bibinfo {pages} {91} (\bibinfo {year} {2005})}\BibitemShut {NoStop}%
\bibitem [{\citenamefont {Bergeret}\ \emph {et~al.}(2005)\citenamefont
  {Bergeret}, \citenamefont {Volkov},\ and\ \citenamefont
  {Efetov}}]{Bergeret2005}%
  \BibitemOpen
  \bibfield  {author} {\bibinfo {author} {\bibfnamefont {F.~S.}\ \bibnamefont
  {Bergeret}}, \bibinfo {author} {\bibfnamefont {A.~F.}\ \bibnamefont
  {Volkov}}, \ and\ \bibinfo {author} {\bibfnamefont {K.~B.}\ \bibnamefont
  {Efetov}},\ }\href@noop {} {\bibfield  {journal} {\bibinfo  {journal} {Rev.
  Mod. Phys.}\ }\textbf {\bibinfo {volume} {77}},\ \bibinfo {pages} {1321}
  (\bibinfo {year} {2005})}\BibitemShut {NoStop}%
\bibitem [{\citenamefont {Tanaka}\ and\ \citenamefont
  {Golubov}(2007)}]{Tanaka2007}%
  \BibitemOpen
  \bibfield  {author} {\bibinfo {author} {\bibfnamefont {Y.}~\bibnamefont
  {Tanaka}}\ and\ \bibinfo {author} {\bibfnamefont {A.~A.}\ \bibnamefont
  {Golubov}},\ }\href {\doibase 10.1103/PhysRevLett.98.037003} {\bibfield
  {journal} {\bibinfo  {journal} {Phys. Rev. Lett.}\ }\textbf {\bibinfo
  {volume} {98}},\ \bibinfo {pages} {037003} (\bibinfo {year}
  {2007})}\BibitemShut {NoStop}%
\bibitem [{\citenamefont {Eschrig}\ and\ \citenamefont
  {L{\"o}fwander}(2008)}]{Eschrig2008}%
  \BibitemOpen
  \bibfield  {author} {\bibinfo {author} {\bibfnamefont {M.}~\bibnamefont
  {Eschrig}}\ and\ \bibinfo {author} {\bibfnamefont {T.}~\bibnamefont
  {L{\"o}fwander}},\ }\href@noop {} {\bibfield  {journal} {\bibinfo  {journal}
  {Nat. Phys.}\ }\textbf {\bibinfo {volume} {4}},\ \bibinfo {pages} {138}
  (\bibinfo {year} {2008})}\BibitemShut {NoStop}%
\bibitem [{\citenamefont {Tanaka}\ \emph {et~al.}(2012)\citenamefont {Tanaka},
  \citenamefont {Sato},\ and\ \citenamefont {Nagaosa}}]{Tanaka2012}%
  \BibitemOpen
  \bibfield  {author} {\bibinfo {author} {\bibfnamefont {Y.}~\bibnamefont
  {Tanaka}}, \bibinfo {author} {\bibfnamefont {M.}~\bibnamefont {Sato}}, \ and\
  \bibinfo {author} {\bibfnamefont {N.}~\bibnamefont {Nagaosa}},\ }\href
  {http://dx.doi.org/10.1143/JPSJ.81.011013} {\bibfield  {journal} {\bibinfo
  {journal} {J. Phys. Soc. Jpn.}\ }\textbf {\bibinfo {volume} {81}},\ \bibinfo
  {pages} {011013} (\bibinfo {year} {2012})}\BibitemShut {NoStop}%
\bibitem [{\citenamefont {Rochalla}\ and\ \citenamefont {Kes~{\rm
  (eds.)}}(2011)}]{Rochalla2011}%
  \BibitemOpen
  \bibfield  {author} {\bibinfo {author} {\bibfnamefont {H.}~\bibnamefont
  {Rochalla}}\ and\ \bibinfo {author} {\bibfnamefont {P.~H.}\ \bibnamefont
  {Kes~{\rm (eds.)}}},\ }\href@noop {} {\emph {\bibinfo {title} {100 Years of
  Superconductivity}}}\ (\bibinfo  {publisher} {CRC Press, Bocca Raton, USA},\
  \bibinfo {year} {2011})\BibitemShut {NoStop}%
\bibitem [{\citenamefont {Solenov}\ \emph {et~al.}(2009)\citenamefont
  {Solenov}, \citenamefont {Martin},\ and\ \citenamefont
  {Mozyrsky}}]{Solenov2009}%
  \BibitemOpen
  \bibfield  {author} {\bibinfo {author} {\bibfnamefont {D.}~\bibnamefont
  {Solenov}}, \bibinfo {author} {\bibfnamefont {I.}~\bibnamefont {Martin}}, \
  and\ \bibinfo {author} {\bibfnamefont {D.}~\bibnamefont {Mozyrsky}},\
  }\href@noop {} {\bibfield  {journal} {\bibinfo  {journal} {Phys. Rev. B}\
  }\textbf {\bibinfo {volume} {79}},\ \bibinfo {pages} {132502} (\bibinfo
  {year} {2009})}\BibitemShut {NoStop}%
\bibitem [{\citenamefont {Kusunose}\ \emph
  {et~al.}(2011{\natexlab{a}})\citenamefont {Kusunose}, \citenamefont
  {Fuseya},\ and\ \citenamefont {Miyake}}]{Kusunose2011}%
  \BibitemOpen
  \bibfield  {author} {\bibinfo {author} {\bibfnamefont {H.}~\bibnamefont
  {Kusunose}}, \bibinfo {author} {\bibfnamefont {Y.}~\bibnamefont {Fuseya}}, \
  and\ \bibinfo {author} {\bibfnamefont {K.}~\bibnamefont {Miyake}},\
  }\href@noop {} {\bibfield  {journal} {\bibinfo  {journal} {J. Phys. Soc.
  Jpn.}\ }\textbf {\bibinfo {volume} {80}},\ \bibinfo {pages} {054702}
  (\bibinfo {year} {2011}{\natexlab{a}})}\BibitemShut {NoStop}%
\bibitem [{\citenamefont {Kusunose}\ \emph
  {et~al.}(2011{\natexlab{b}})\citenamefont {Kusunose}, \citenamefont
  {Fuseya},\ and\ \citenamefont {Miyake}}]{Kusunose2011a}%
  \BibitemOpen
  \bibfield  {author} {\bibinfo {author} {\bibfnamefont {H.}~\bibnamefont
  {Kusunose}}, \bibinfo {author} {\bibfnamefont {Y.}~\bibnamefont {Fuseya}}, \
  and\ \bibinfo {author} {\bibfnamefont {K.}~\bibnamefont {Miyake}},\
  }\href@noop {} {\bibfield  {journal} {\bibinfo  {journal} {J. Phys. Soc.
  Jpn.}\ }\textbf {\bibinfo {volume} {80}},\ \bibinfo {pages} {044711}
  (\bibinfo {year} {2011}{\natexlab{b}})}\BibitemShut {NoStop}%
\bibitem [{\citenamefont {Fuseya}\ \emph {et~al.}(2003)\citenamefont {Fuseya},
  \citenamefont {Kohno},\ and\ \citenamefont {Miyake}}]{Fuseya2003}%
  \BibitemOpen
  \bibfield  {author} {\bibinfo {author} {\bibfnamefont {Y.}~\bibnamefont
  {Fuseya}}, \bibinfo {author} {\bibfnamefont {H.}~\bibnamefont {Kohno}}, \
  and\ \bibinfo {author} {\bibfnamefont {K.}~\bibnamefont {Miyake}},\
  }\href@noop {} {\bibfield  {journal} {\bibinfo  {journal} {J. Phys. Soc.
  Jpn.}\ }\textbf {\bibinfo {volume} {72}},\ \bibinfo {pages} {2914} (\bibinfo
  {year} {2003})}\BibitemShut {NoStop}%
\bibitem [{\citenamefont {Matsumoto}\ \emph {et~al.}(2012)\citenamefont
  {Matsumoto}, \citenamefont {Koga},\ and\ \citenamefont
  {Kusunose}}]{Matsumoto2012}%
  \BibitemOpen
  \bibfield  {author} {\bibinfo {author} {\bibfnamefont {M.}~\bibnamefont
  {Matsumoto}}, \bibinfo {author} {\bibfnamefont {M.}~\bibnamefont {Koga}}, \
  and\ \bibinfo {author} {\bibfnamefont {H.}~\bibnamefont {Kusunose}},\
  }\href@noop {} {\bibfield  {journal} {\bibinfo  {journal} {J. Phys. Soc.
  Jpn.}\ }\textbf {\bibinfo {volume} {81}},\ \bibinfo {pages} {033702}
  (\bibinfo {year} {2012})}\BibitemShut {NoStop}%
\bibitem [{\citenamefont {Black-Schaffer}\ and\ \citenamefont
  {Balatsky}(2013)}]{Black-Schaffer2013}%
  \BibitemOpen
  \bibfield  {author} {\bibinfo {author} {\bibfnamefont {A.~M.}\ \bibnamefont
  {Black-Schaffer}}\ and\ \bibinfo {author} {\bibfnamefont {A.~V.}\
  \bibnamefont {Balatsky}},\ }\href@noop {} {\bibfield  {journal} {\bibinfo
  {journal} {Phys. Rev. B}\ }\textbf {\bibinfo {volume} {88}},\ \bibinfo
  {pages} {104514} (\bibinfo {year} {2013})}\BibitemShut {NoStop}%
\bibitem [{\citenamefont {Nagamatsu}\ \emph {et~al.}(2001)\citenamefont
  {Nagamatsu}, \citenamefont {Nakagawa}, \citenamefont {Muranaka},
  \citenamefont {Zenitani},\ and\ \citenamefont {Akimitsu}}]{Nagamatsu2001}%
  \BibitemOpen
  \bibfield  {author} {\bibinfo {author} {\bibfnamefont {J.}~\bibnamefont
  {Nagamatsu}}, \bibinfo {author} {\bibfnamefont {N.}~\bibnamefont {Nakagawa}},
  \bibinfo {author} {\bibfnamefont {T.}~\bibnamefont {Muranaka}}, \bibinfo
  {author} {\bibfnamefont {Y.}~\bibnamefont {Zenitani}}, \ and\ \bibinfo
  {author} {\bibfnamefont {J.}~\bibnamefont {Akimitsu}},\ }\href@noop {}
  {\bibfield  {journal} {\bibinfo  {journal} {Nature}\ }\textbf {\bibinfo
  {volume} {410}},\ \bibinfo {pages} {63} (\bibinfo {year} {2001})}\BibitemShut
  {NoStop}%
\bibitem [{\citenamefont {Souma}\ \emph {et~al.}(2003)\citenamefont {Souma},
  \citenamefont {Machida}, \citenamefont {Sato}, \citenamefont {Takahashi},
  \citenamefont {Matsui}, \citenamefont {Wang}, \citenamefont {Ding},
  \citenamefont {Kaminski}, \citenamefont {Campuzano}, \citenamefont {Sasaki},\
  and\ \citenamefont {Kadowaki}}]{Souma2003}%
  \BibitemOpen
  \bibfield  {author} {\bibinfo {author} {\bibfnamefont {S.}~\bibnamefont
  {Souma}}, \bibinfo {author} {\bibfnamefont {Y.}~\bibnamefont {Machida}},
  \bibinfo {author} {\bibfnamefont {T.}~\bibnamefont {Sato}}, \bibinfo {author}
  {\bibfnamefont {T.}~\bibnamefont {Takahashi}}, \bibinfo {author}
  {\bibfnamefont {H.}~\bibnamefont {Matsui}}, \bibinfo {author} {\bibfnamefont
  {S.-C.}\ \bibnamefont {Wang}}, \bibinfo {author} {\bibfnamefont
  {H.}~\bibnamefont {Ding}}, \bibinfo {author} {\bibfnamefont {A.}~\bibnamefont
  {Kaminski}}, \bibinfo {author} {\bibfnamefont {J.~C.}\ \bibnamefont
  {Campuzano}}, \bibinfo {author} {\bibfnamefont {S.}~\bibnamefont {Sasaki}}, \
  and\ \bibinfo {author} {\bibfnamefont {K.}~\bibnamefont {Kadowaki}},\
  }\href@noop {} {\bibfield  {journal} {\bibinfo  {journal} {Nature}\ }\textbf
  {\bibinfo {volume} {423}},\ \bibinfo {pages} {65} (\bibinfo {year}
  {2003})}\BibitemShut {NoStop}%
\bibitem [{\citenamefont {An}\ and\ \citenamefont {Pickett}(2001)}]{An2001}%
  \BibitemOpen
  \bibfield  {author} {\bibinfo {author} {\bibfnamefont {J.~M.}\ \bibnamefont
  {An}}\ and\ \bibinfo {author} {\bibfnamefont {W.~E.}\ \bibnamefont
  {Pickett}},\ }\href {\doibase 10.1103/PhysRevLett.86.4366} {\bibfield
  {journal} {\bibinfo  {journal} {Phys. Rev. Lett.}\ }\textbf {\bibinfo
  {volume} {86}},\ \bibinfo {pages} {4366} (\bibinfo {year}
  {2001})}\BibitemShut {NoStop}%
\bibitem [{\citenamefont {Kortus}\ \emph {et~al.}(2001)\citenamefont {Kortus},
  \citenamefont {Mazin}, \citenamefont {Belashchenko}, \citenamefont
  {Antropov},\ and\ \citenamefont {Boyer}}]{Kortus2001}%
  \BibitemOpen
  \bibfield  {author} {\bibinfo {author} {\bibfnamefont {J.}~\bibnamefont
  {Kortus}}, \bibinfo {author} {\bibfnamefont {I.~I.}\ \bibnamefont {Mazin}},
  \bibinfo {author} {\bibfnamefont {K.~D.}\ \bibnamefont {Belashchenko}},
  \bibinfo {author} {\bibfnamefont {V.~P.}\ \bibnamefont {Antropov}}, \ and\
  \bibinfo {author} {\bibfnamefont {L.~L.}\ \bibnamefont {Boyer}},\ }\href
  {\doibase 10.1103/PhysRevLett.86.4656} {\bibfield  {journal} {\bibinfo
  {journal} {Phys. Rev. Lett.}\ }\textbf {\bibinfo {volume} {86}},\ \bibinfo
  {pages} {4656} (\bibinfo {year} {2001})}\BibitemShut {NoStop}%
\bibitem [{\citenamefont {Bohnen}\ \emph {et~al.}(2001)\citenamefont {Bohnen},
  \citenamefont {Heid},\ and\ \citenamefont {Renker}}]{Bohnen2001}%
  \BibitemOpen
  \bibfield  {author} {\bibinfo {author} {\bibfnamefont {K.-P.}\ \bibnamefont
  {Bohnen}}, \bibinfo {author} {\bibfnamefont {R.}~\bibnamefont {Heid}}, \ and\
  \bibinfo {author} {\bibfnamefont {B.}~\bibnamefont {Renker}},\ }\href
  {\doibase 10.1103/PhysRevLett.86.5771} {\bibfield  {journal} {\bibinfo
  {journal} {Phys. Rev. Lett.}\ }\textbf {\bibinfo {volume} {86}},\ \bibinfo
  {pages} {5771} (\bibinfo {year} {2001})}\BibitemShut {NoStop}%
\bibitem [{\citenamefont {Choi}\ \emph
  {et~al.}(2002{\natexlab{a}})\citenamefont {Choi}, \citenamefont {Roundy},
  \citenamefont {Sun}, \citenamefont {Cohen},\ and\ \citenamefont
  {Louie}}]{Choi2002a}%
  \BibitemOpen
  \bibfield  {author} {\bibinfo {author} {\bibfnamefont {H.~J.}\ \bibnamefont
  {Choi}}, \bibinfo {author} {\bibfnamefont {D.}~\bibnamefont {Roundy}},
  \bibinfo {author} {\bibfnamefont {H.}~\bibnamefont {Sun}}, \bibinfo {author}
  {\bibfnamefont {M.~L.}\ \bibnamefont {Cohen}}, \ and\ \bibinfo {author}
  {\bibfnamefont {S.~G.}\ \bibnamefont {Louie}},\ }\href@noop {} {\bibfield
  {journal} {\bibinfo  {journal} {Phys. Rev. B}\ }\textbf {\bibinfo {volume}
  {66}},\ \bibinfo {pages} {020513} (\bibinfo {year}
  {2002}{\natexlab{a}})}\BibitemShut {NoStop}%
\bibitem [{\citenamefont {Szab\'o}\ \emph {et~al.}(2001)\citenamefont
  {Szab\'o}, \citenamefont {Samuely}, \citenamefont {Ka\v{c}mar\v{c}\'ik},
  \citenamefont {Klein}, \citenamefont {Marcus}, \citenamefont {Fruchart},
  \citenamefont {Miraglia}, \citenamefont {Marcenat},\ and\ \citenamefont
  {Jansen}}]{Szabo2001}%
  \BibitemOpen
  \bibfield  {author} {\bibinfo {author} {\bibfnamefont {P.}~\bibnamefont
  {Szab\'o}}, \bibinfo {author} {\bibfnamefont {P.}~\bibnamefont {Samuely}},
  \bibinfo {author} {\bibfnamefont {J.}~\bibnamefont {Ka\v{c}mar\v{c}\'ik}},
  \bibinfo {author} {\bibfnamefont {T.}~\bibnamefont {Klein}}, \bibinfo
  {author} {\bibfnamefont {J.}~\bibnamefont {Marcus}}, \bibinfo {author}
  {\bibfnamefont {D.}~\bibnamefont {Fruchart}}, \bibinfo {author}
  {\bibfnamefont {S.}~\bibnamefont {Miraglia}}, \bibinfo {author}
  {\bibfnamefont {C.}~\bibnamefont {Marcenat}}, \ and\ \bibinfo {author}
  {\bibfnamefont {A.~G.~M.}\ \bibnamefont {Jansen}},\ }\href {\doibase
  10.1103/PhysRevLett.87.137005} {\bibfield  {journal} {\bibinfo  {journal}
  {Phys. Rev. Lett.}\ }\textbf {\bibinfo {volume} {87}},\ \bibinfo {pages}
  {137005} (\bibinfo {year} {2001})}\BibitemShut {NoStop}%
\bibitem [{\citenamefont {Choi}\ \emph
  {et~al.}(2002{\natexlab{b}})\citenamefont {Choi}, \citenamefont {Roundy},
  \citenamefont {Sun}, \citenamefont {Cohen},\ and\ \citenamefont
  {Louie}}]{Choi2002}%
  \BibitemOpen
  \bibfield  {author} {\bibinfo {author} {\bibfnamefont {H.~J.}\ \bibnamefont
  {Choi}}, \bibinfo {author} {\bibfnamefont {D.}~\bibnamefont {Roundy}},
  \bibinfo {author} {\bibfnamefont {H.}~\bibnamefont {Sun}}, \bibinfo {author}
  {\bibfnamefont {M.~L.}\ \bibnamefont {Cohen}}, \ and\ \bibinfo {author}
  {\bibfnamefont {S.~G.}\ \bibnamefont {Louie}},\ }\href@noop {} {\bibfield
  {journal} {\bibinfo  {journal} {Nature}\ }\textbf {\bibinfo {volume} {418}},\
  \bibinfo {pages} {758} (\bibinfo {year} {2002}{\natexlab{b}})}\BibitemShut
  {NoStop}%
\bibitem [{\citenamefont {Chen}\ \emph {et~al.}(2012)\citenamefont {Chen},
  \citenamefont {Dai}, \citenamefont {Zhuang}, \citenamefont {Li},
  \citenamefont {Carabello}, \citenamefont {Lambert}, \citenamefont {Mlack},
  \citenamefont {Ramos},\ and\ \citenamefont {Xi}}]{Chen2012}%
  \BibitemOpen
  \bibfield  {author} {\bibinfo {author} {\bibfnamefont {K.}~\bibnamefont
  {Chen}}, \bibinfo {author} {\bibfnamefont {W.}~\bibnamefont {Dai}}, \bibinfo
  {author} {\bibfnamefont {C.}~\bibnamefont {Zhuang}}, \bibinfo {author}
  {\bibfnamefont {Q.}~\bibnamefont {Li}}, \bibinfo {author} {\bibfnamefont
  {S.}~\bibnamefont {Carabello}}, \bibinfo {author} {\bibfnamefont {J.~G.}\
  \bibnamefont {Lambert}}, \bibinfo {author} {\bibfnamefont {J.~T.}\
  \bibnamefont {Mlack}}, \bibinfo {author} {\bibfnamefont {R.~C.}\ \bibnamefont
  {Ramos}}, \ and\ \bibinfo {author} {\bibfnamefont {X.~X.}\ \bibnamefont
  {Xi}},\ }\href {http://dx.doi.org/10.1038/ncomms1626} {\bibfield  {journal}
  {\bibinfo  {journal} {Nat. Commun.}\ }\textbf {\bibinfo {volume} {3}},\
  \bibinfo {pages} {619} (\bibinfo {year} {2012})}\BibitemShut {NoStop}%
\bibitem [{\citenamefont {Eom}\ \emph {et~al.}(2001)\citenamefont {Eom},
  \citenamefont {Lee}, \citenamefont {Choi}, \citenamefont {Belenky},
  \citenamefont {Song}, \citenamefont {Cooley}, \citenamefont {Naus},
  \citenamefont {Patnaik}, \citenamefont {Jiang}, \citenamefont {Rikel},
  \citenamefont {Polyanskii}, \citenamefont {Gurevich}, \citenamefont {Cai},
  \citenamefont {Bu}, \citenamefont {Babcock}, \citenamefont {Hellstrom},
  \citenamefont {Larbalestier}, \citenamefont {Rogado}, \citenamefont {Regan},
  \citenamefont {Hayward}, \citenamefont {He}, \citenamefont {Slusky},
  \citenamefont {Inumaru}, \citenamefont {Haas},\ and\ \citenamefont
  {Cava}}]{Eom2001}%
  \BibitemOpen
  \bibfield  {author} {\bibinfo {author} {\bibfnamefont {C.~B.}\ \bibnamefont
  {Eom}}, \bibinfo {author} {\bibfnamefont {M.~K.}\ \bibnamefont {Lee}},
  \bibinfo {author} {\bibfnamefont {J.~H.}\ \bibnamefont {Choi}}, \bibinfo
  {author} {\bibfnamefont {L.~J.}\ \bibnamefont {Belenky}}, \bibinfo {author}
  {\bibfnamefont {X.}~\bibnamefont {Song}}, \bibinfo {author} {\bibfnamefont
  {L.~D.}\ \bibnamefont {Cooley}}, \bibinfo {author} {\bibfnamefont {M.~T.}\
  \bibnamefont {Naus}}, \bibinfo {author} {\bibfnamefont {S.}~\bibnamefont
  {Patnaik}}, \bibinfo {author} {\bibfnamefont {J.}~\bibnamefont {Jiang}},
  \bibinfo {author} {\bibfnamefont {M.}~\bibnamefont {Rikel}}, \bibinfo
  {author} {\bibfnamefont {A.}~\bibnamefont {Polyanskii}}, \bibinfo {author}
  {\bibfnamefont {A.}~\bibnamefont {Gurevich}}, \bibinfo {author}
  {\bibfnamefont {X.~Y.}\ \bibnamefont {Cai}}, \bibinfo {author} {\bibfnamefont
  {S.~D.}\ \bibnamefont {Bu}}, \bibinfo {author} {\bibfnamefont {S.~E.}\
  \bibnamefont {Babcock}}, \bibinfo {author} {\bibfnamefont {E.~E.}\
  \bibnamefont {Hellstrom}}, \bibinfo {author} {\bibfnamefont {D.~C.}\
  \bibnamefont {Larbalestier}}, \bibinfo {author} {\bibfnamefont
  {N.}~\bibnamefont {Rogado}}, \bibinfo {author} {\bibfnamefont {K.~A.}\
  \bibnamefont {Regan}}, \bibinfo {author} {\bibfnamefont {M.~A.}\ \bibnamefont
  {Hayward}}, \bibinfo {author} {\bibfnamefont {T.}~\bibnamefont {He}},
  \bibinfo {author} {\bibfnamefont {J.~S.}\ \bibnamefont {Slusky}}, \bibinfo
  {author} {\bibfnamefont {K.}~\bibnamefont {Inumaru}}, \bibinfo {author}
  {\bibfnamefont {M.~K.}\ \bibnamefont {Haas}}, \ and\ \bibinfo {author}
  {\bibfnamefont {R.~J.}\ \bibnamefont {Cava}},\ }\href
  {http://dx.doi.org/10.1038/35079018} {\bibfield  {journal} {\bibinfo
  {journal} {Nature}\ }\textbf {\bibinfo {volume} {411}},\ \bibinfo {pages}
  {558} (\bibinfo {year} {2001})}\BibitemShut {NoStop}%
\bibitem [{\citenamefont {Gurevich}(2007)}]{Gurevich2007}%
  \BibitemOpen
  \bibfield  {author} {\bibinfo {author} {\bibfnamefont {A.}~\bibnamefont
  {Gurevich}},\ }\href {\doibase http://dx.doi.org/10.1016/j.physc.2007.01.008}
  {\bibfield  {journal} {\bibinfo  {journal} {Physica C}\ }\textbf {\bibinfo
  {volume} {456}},\ \bibinfo {pages} {160 } (\bibinfo {year}
  {2007})}\BibitemShut {NoStop}%
\end{thebibliography}
\end{document}